\documentclass[aps,prl,twocolumn,floatfix,showpacs,superscriptaddress]{revtex4-1}
\usepackage{graphicx,amsmath,amsfonts,mathrsfs}
\usepackage{longtable}
\usepackage{wasysym}
\usepackage{trsym}
\usepackage{marvosym}

\begin{document}

\title{Quantum spin liquid and cluster Mott insulator phases in the Mo$_{3}$O$_{8}$ magnets}

\author{S.~A.~Nikolaev}
\email{nikolaev.s.aa@m.titech.ac.jp}
\email{saishi@inbox.ru}
\affiliation{Institute of Innovative Research, Tokyo Institute of Technology, 4259 Nagatsuta, Midori, Yokohama 226-8503, Japan}
\affiliation{International Center for Materials Nanoarchitectonics, National Institute for Materials Science, 1-1 Namiki, Tsukuba, Ibaraki 305-0044, Japan}
\author{I.~V.~Solovyev}
\affiliation{International Center for Materials Nanoarchitectonics, National Institute for Materials Science, 1-1 Namiki, Tsukuba, Ibaraki 305-0044, Japan}
\affiliation{Department of Theoretical Physics and Applied Mathematics, Ural Federal University, Mira St. 19, 620002 Yekaterinburg, Russia}
\author{S.~V.~Streltsov}
\affiliation{Institute of Metal Physics, S. Kovalevskoy Street 18, 620108 Yekaterinburg, Russia}
\affiliation{Department of Theoretical Physics and Applied Mathematics, Ural Federal University, Mira St. 19, 620002 Yekaterinburg, Russia}
\date{\today}

\begin{abstract}
We unveil the microscopic origin of largely debated magnetism in the Mo$_{3}$O$_{8}$ cluster systems. Upon considering an extended Hubbard model at 1/6 filling on the anisotropic kagom\'e lattice formed by the Mo atoms, we argue that its ground state is determined by the competition between kinetic energy and intersite Coulomb interactions, which is controlled by the trimerisation of the kagom\'e lattice into the Mo$_{3}$O$_{13}$ clusters. Based on first-principles calculations, we show that the strong interaction limit is realised in LiZn$_{2}$Mo$_{3}$O$_{8}$ revealing a plaquette charge order with unpaired spins at the resonating hexagons, whose origin is solely related to the opposite signs of intracluster and intercluster hoppings, in contrast to all previous scenarios. On the other hand, both Li$_{2}$InMo$_{3}$O$_{8}$ and Li$_{2}$ScMo$_{3}$O$_{8}$ are demonstrated to fall into the weak interaction limit where the electrons are well localised at the Mo$_{3}$O$_{13}$ clusters. While the former is found to exhibit long-range antiferromagnetic order, the latter is more likely to reveal short-range order with quantum spin liquid-like excitations. Our results not only reproduce most of the experimentally observed features of these unique materials, but will also help to describe various properties in other quantum cluster magnets.
\end{abstract}
\maketitle

\par \emph{Introduction}. Geometrically frustrated quantum systems lie at the core of research activity revolving around a putative quantum spin liquid (QSL) state that displays long-range quantum entanglement, charge fractionalisation, and emergent gauge structures~\cite{moessner,patrick1,balents,review}. Of particular importance are spin models on the triangular and kagom\'e lattices featuring various types of QSL~\cite{spinliq1,spinliq2,spinliq3,spinliq4,spinliq5}, whose material realisation has been an ongoing endeavour in condensed matter physics with only a few reasonable candidates proposed so far~\cite{vorbot,herb1,herb2,bedt,dmit1,dmit2}.

%%%%%%%%%%%%%%%%%%%
\begin{figure}[b!]
\begin{center}
\includegraphics[width=0.46\textwidth]{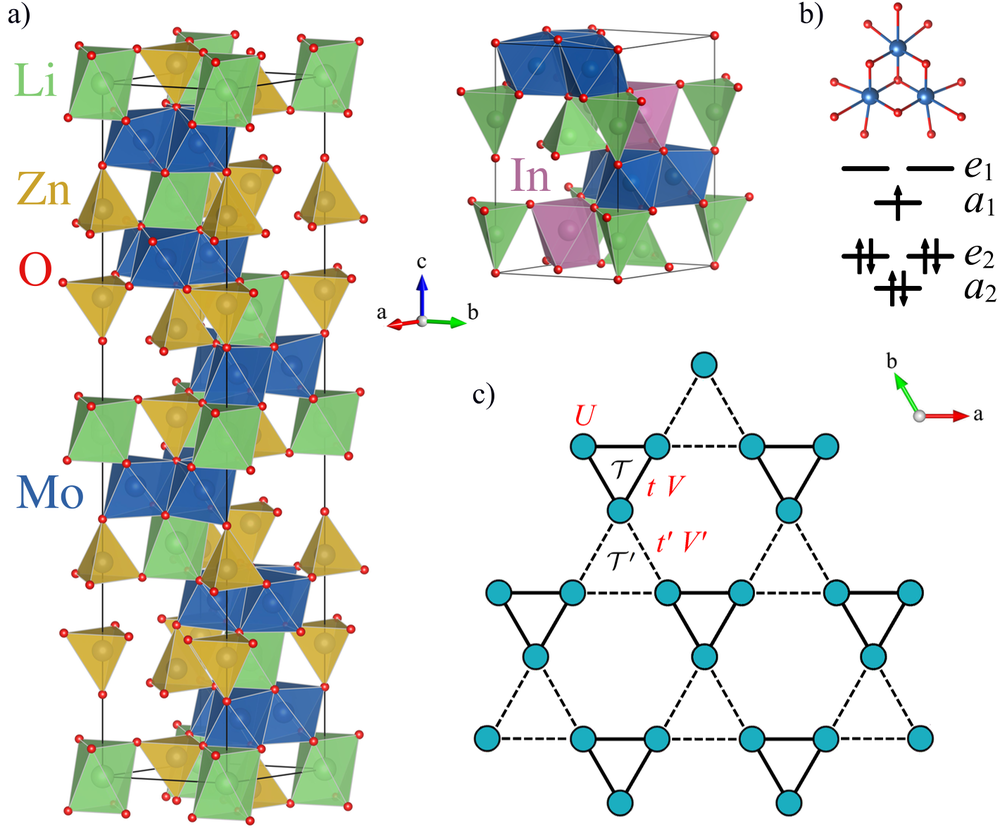}
\end{center}
\caption{a) Crystal structures of LiZn$_{2}$Mo$_{3}$O$_{8}$ and Li$_{2}$InMo$_{3}$O$_{8}$, visualised with \texttt{VESTA}~\cite{vesta}. Li$_{2}$ScMo$_{3}$O$_{8}$ is isostructural to Li$_{2}$InMo$_{3}$O$_{8}$;~b) Molecular levels of the Mo$_{3}$O$_{13}$ cluster filled with seven electrons;~c) Schematics of the extended Hubbard model on the anisotropic kagom\'e lattice formed by the Mo sites. Nonequivalent ``up" and ``down" triangles are denoted as $\mathcal{T}'$ and $\mathcal{T}$, respectively.}
\label{fig:kagome}
\end{figure}
%%%%%%%%%%%%%%%%%%%

\par During the past few years, the Mo$_{3}$O$_{13}$ cluster magnets have attracted a great deal of both experimental and theoretical attention as a new candidate to host QSL. In these compounds, the Mo atoms arranged in anisotropic kagom\'e layers are trimerised, and the [Mo$_{3}$O$_{13}$]$^{15-}$ clusters form a triangular lattice, as shown in Fig.~\ref{fig:kagome}a~\cite{carroll,cotton}. As sketched in Fig.~\ref{fig:kagome}b, six out of seven valence electrons in the Mo$_{3}$O$_{13}$ cluster are responsible for a strong intracluster metal-metal bonding, and the seventh electron remains unpaired occupying a totally symmetric molecular $a_{1}$ state.
\par LiZn$_{2}$Mo$_{3}$O$_{8}$ was first reported to exhibit a QSL behaviour~\cite{sheckel1,sheckel2,sheckel3}.~The magnetic susceptibility of LiZn$_{2}$Mo$_{3}$O$_{8}$ has been experimentally shown to follow a Curie-Weiss law with low- and high-temperature regimes transitioning at 96~K, whose Curie constants are related as $C_{L}\approx C_{H}/3$ and where the disappearance of 2/3 of paramagnetic spins was attributed to valence bond condensation on the triangular lattice of the Mo$_{3}$O$_{13}$ clusters. In a first attempt to explain these unusual features, the authors of Ref.~\cite{flint} suggested the formation of an emergent honeycomb lattice due to opposite rotations of the Mo$_{3}$O$_{13}$ clusters effectively decoupling the central cluster with an orphan paramagnetic spin. Another scenario was outlined in Ref.~\cite{chen}, where a plaquette charge order (PCO) existing in a Mott insulator on the anisotropic kagom\'e lattice at 1/6 filling was conjectured to host a $U(1)$ quantum spin liquid state with the spinon Fermi surface that is reconstructed at low temperatures filling 2/3 of the spinon states.  
\par However, the adequacy of the proposed mechanisms was questioned with recently synthesised Li$_{2}$InMo$_{3}$O$_{8}$ and Li$_{2}$ScMo$_{3}$O$_{8}$, both featuring magnetic moments well localised at the Mo$_{3}$O$_{13}$ clusters. While the former was identified with a Ne\'el 120$^\circ$ magnetic order at $T_{N}=12$~K, for the latter no magnetic ordering has been observed down to 0.5~K~\cite{haraguchi1,haraguchi2}. Instead, muon spin rotation and inelastic neutron scattering measurements suggested that Li$_{2}$ScMo$_{3}$O$_{8}$ undergoes a short-range magnetic order below 4~K with QSL-like excitations. 
\par Despite having similar crystal structures, these systems manifest essentially unalike magnetic properties, whose enigmatic origin remains an unsolved problem. In this Letter, upon revising a single-orbital extended Hubbard model on the anisotropic kagom\'e lattice at 1/6 filling, we uncover novel regimes of the plaquette charge ordered and cluster Mott insulator states governed by the interplay of kinetic energy and intersite Coulomb interactions, that were overlooked in previous studies~\cite{flint,chen,chen2}. By means of first-principles calculations, we will demonstrate that their appearance is related to the formation of the Mo$_{3}$O$_{13}$ clusters and these states indeed realise in LiZn$_{2}$Mo$_{3}$O$_{8}$, Li$_{2}$ScMo$_{3}$O$_{8}$, and Li$_{2}$InMo$_{3}$O$_{8}$.
\par The model of interest shown in Fig.~\ref{fig:kagome}c with one electron per $\mathcal{T}$ triangle reads:
\begin{equation}
\begin{aligned}
\mathcal{H}&=\sum_{\substack{\langle mm' \rangle \in \mathcal{T} \\ \sigma}} t\big(c^{\dagger\sigma}_{m}c^{\,\sigma}_{m'} + \mathrm{H.c.}\big) +Vn^{}_{m}n^{}_{m'} + U\sum_{m}n_{m}^{\uparrow}n_{m}^{\downarrow} \\
&+\sum_{\substack{\langle mm' \rangle \in \mathcal{T}'\\ \sigma'}} t'\big(c^{\dagger\sigma}_{m}c^{\,\sigma}_{m'} + \mathrm{H.c.}\big) +V'n^{}_{m}n^{}_{m'}, 
\end{aligned}
\label{eq:hubsingle}
\end{equation}
\noindent where $c^{\dagger\sigma}_{m}$ ($c^{\phantom{\dagger}\sigma}_{m}$) creates (annihilates) an electron with spin $\sigma$ at site $m$, $n^{\sigma}_{m}=c^{\dagger\sigma}_{m}c^{\phantom{\dagger}\sigma}_{m}$ is the density operator ($n_{m}^{}=n^{\uparrow}_{m}+n^{\downarrow}_{m}$), $t$ and $t'$ stand for intracluster and intercluster hopping parameters defined on the $\mathcal{T}$ and $\mathcal{T}'$ triangles, respectively, and the interaction terms include the on-site $U$, intracluster $V$, and intercluster $V'$ Coulomb repulsions. Taking a shorter bond length in $\mathcal{T}$, we assume that $V'<V\ll U$ and $|t'|<|t|$, and for the reasons shown below we enforce electron localisation at the $\mathcal{T}$ triangles by taking $t<0$ and $t'>0$.
\par
\begin{figure}[t]
\begin{center}
\includegraphics[width=0.48\textwidth]{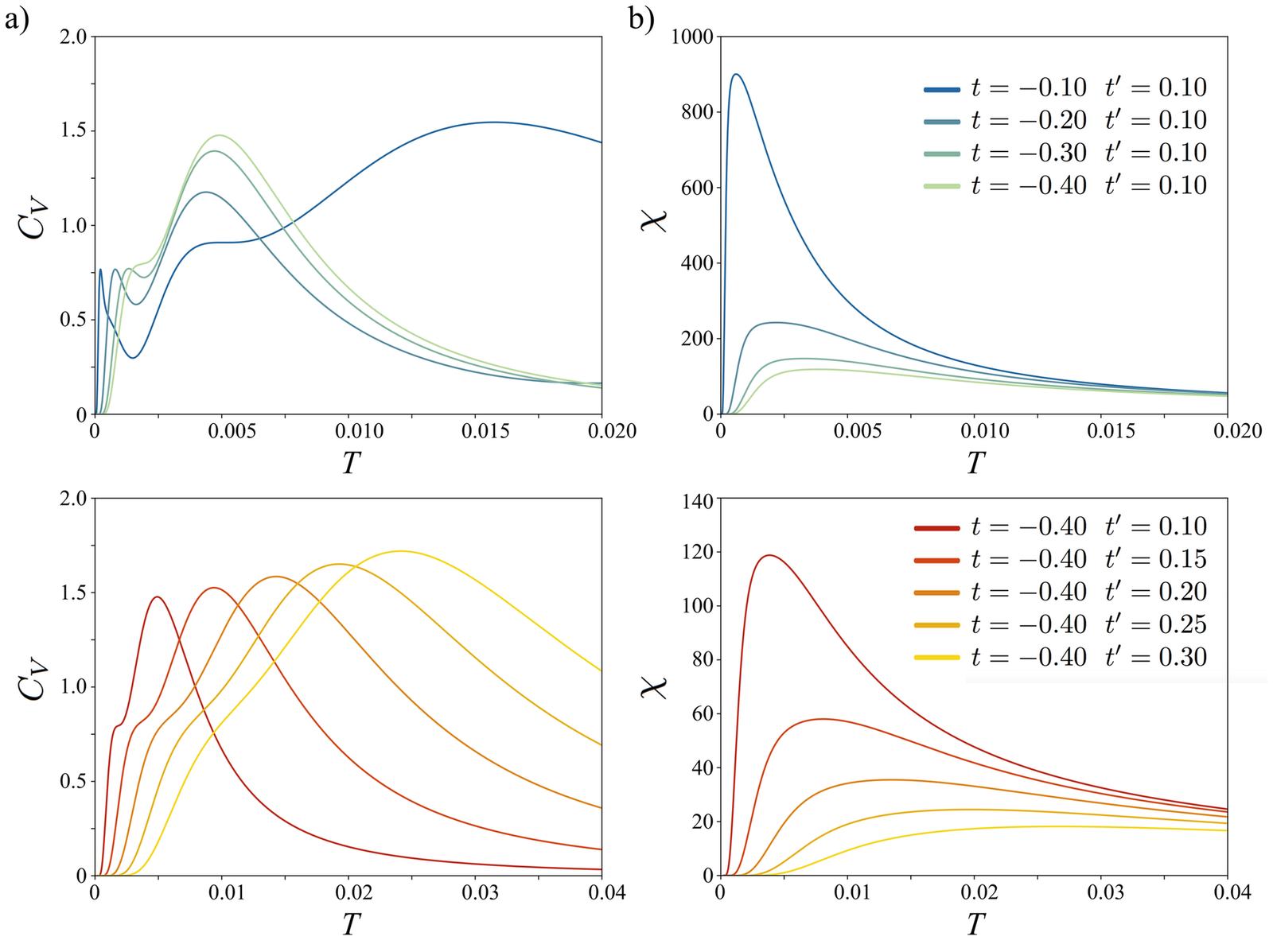}
\end{center}
\caption{Specific heat (a) and spin susceptibility (b) of the extended Hubbard model, Eq.~(\ref{eq:hubsingle}), at 1/6 filling calculated with exact diagonalisation on a $2\times2$ supercell with 12 sites and periodic boundary conditions as a function of $t$ and $t'$ with $U=2.0$, $V=1.0$, and $V'=0.8$.}
\label{fig:ed}
\end{figure}

 The results of exact diagonalisation for Eq.~(\ref{eq:hubsingle}) at 1/6 filling are shown in Fig.~\ref{fig:ed}. Specific heat has an evident instability when $t/V'$ and $t'/V$ are small, while as they increase the system can develop long-range magnetic order. Thus, one can see that there are several regimes depending on the values of $t/V'$ and $t'/V$, and below we will address two different limits of Eq.~(\ref{eq:hubsingle}).

%%%%%%%%%%%%%%%%%%%
%%%%%%%%%%%%%%%%%%%

%%%%%%%%%%%%%%%%%%%
\begin{figure}[b]
\begin{center}
\includegraphics[width=0.48\textwidth]{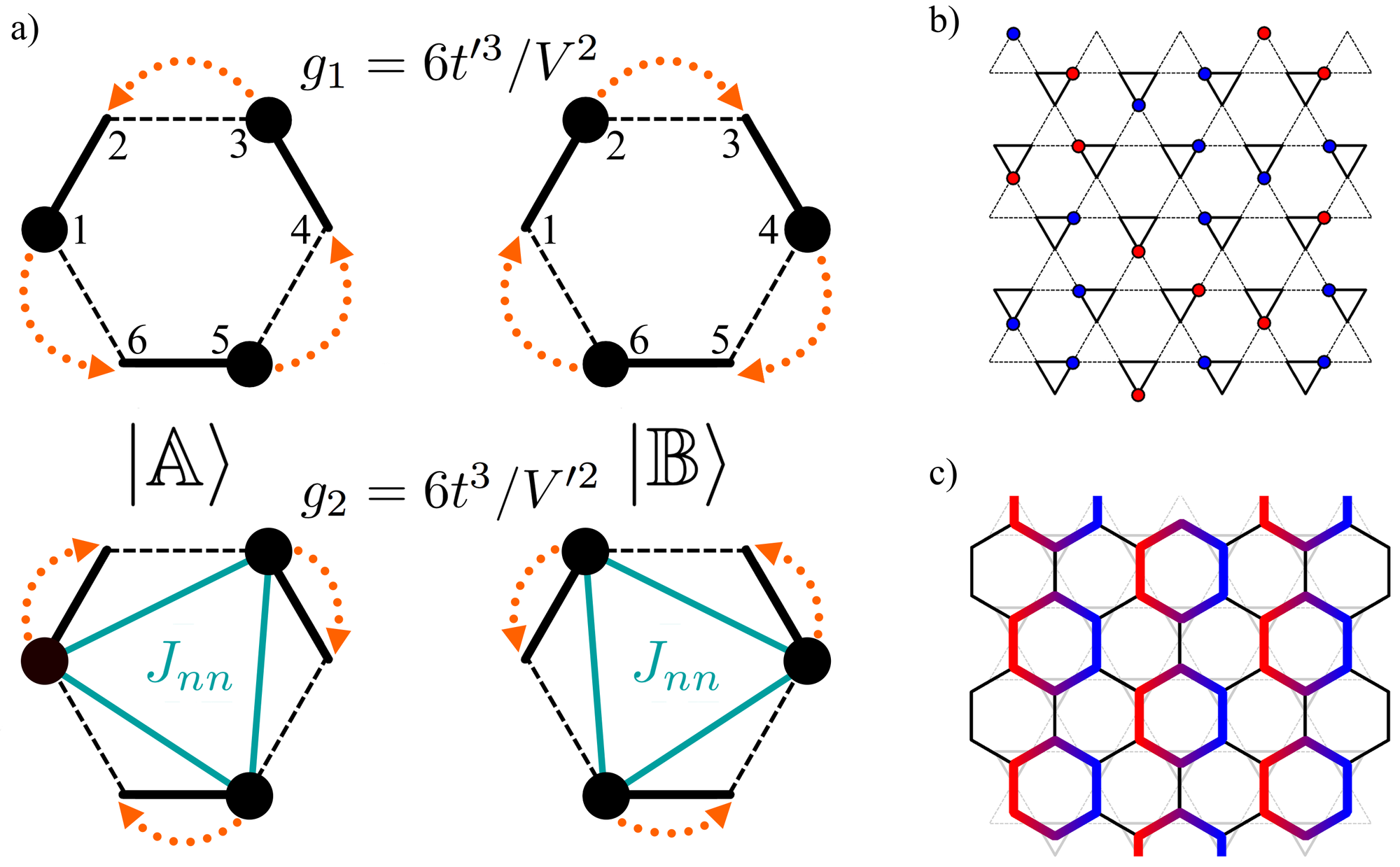}
\end{center}
\caption{a) Ring tunnelling processes in the hexagon; b) Charge order in the strong interaction limit (blue and red circles stand for the spin-up and spin-down electrons, respectively) b) Plaquette charge ordered phase on the dual hexagonal lattice.}
\label{fig:hexagon}
\end{figure}
%%%%%%%%%%%%%%%%%%%

\par \emph{Plaquette charge order}. Let us consider $t \ll V'$ and $t'\ll V$. Due to 1/6 filling, the Hubbard $U$ cannot localise electrons on the lattice sites, and as a result they move without encountering any double occupancy. Since $U$ is not operative, it is the intersite $V$ and $V'$ that are responsible for electron localisation leading to a highly degenerate charge ordered state, where each corner-sharing triangle hosts exactly one electron. This degeneracy is further lifted by hopping parameters that induce collective tunnelling processes, when the electrons hop either clockwise or counter-clockwise along the $\mathcal{T}$ and $\mathcal{T'}$ bonds stabilising a charge pattern with three electrons at the hexagons, as shown in Fig.~\ref{fig:hexagon}a and ~\ref{fig:hexagon}b. To lowest order in $t/V'$ and $t'/V$, it corresponds to the quantum dimer model for two plaquette states $|\mathbb{A}\rangle = c^{\dagger\sigma}_{5}c^{\dagger\sigma'}_{3}c^{\dagger\sigma''}_{1}|0\rangle$ and $|\mathbb{B}\rangle = c^{\dagger\sigma}_{6}c^{\dagger\sigma'}_{4}c^{\dagger\sigma''}_{2}|0\rangle$, $\mathcal{H}_{\hexagon}=\sum_{\hexagon}\sum_{\sigma\sigma'\sigma''}(g_{1}+g_{2})\big(|\mathbb{A}\rangle\langle\mathbb{B}|+|\mathbb{B}\rangle\langle\mathbb{A}|\big)$ with $g_{1}=6t'^{3}/V^{2}$  and $g_{2}=6t^{3}/V'^{2}$, where the sum runs over all hexagons~\cite{dimer1,dimer2}. When mapped onto the dual hexagonal lattice, the ground state of $\mathcal{H}_{\hexagon}$ for spinless electrons is described by the PCO shown in Fig.~\ref{fig:hexagon}c with an emergent triangular lattice of resonating hexagons, that will be regarded as the strong interaction limit of Eq.~(\ref{eq:hubsingle})~\cite{kinetic,nishimoto,brien}.

%%%%%%%%%%%%%%%%%%%
\begin{figure}[b]
\begin{center}
\includegraphics[width=0.48\textwidth]{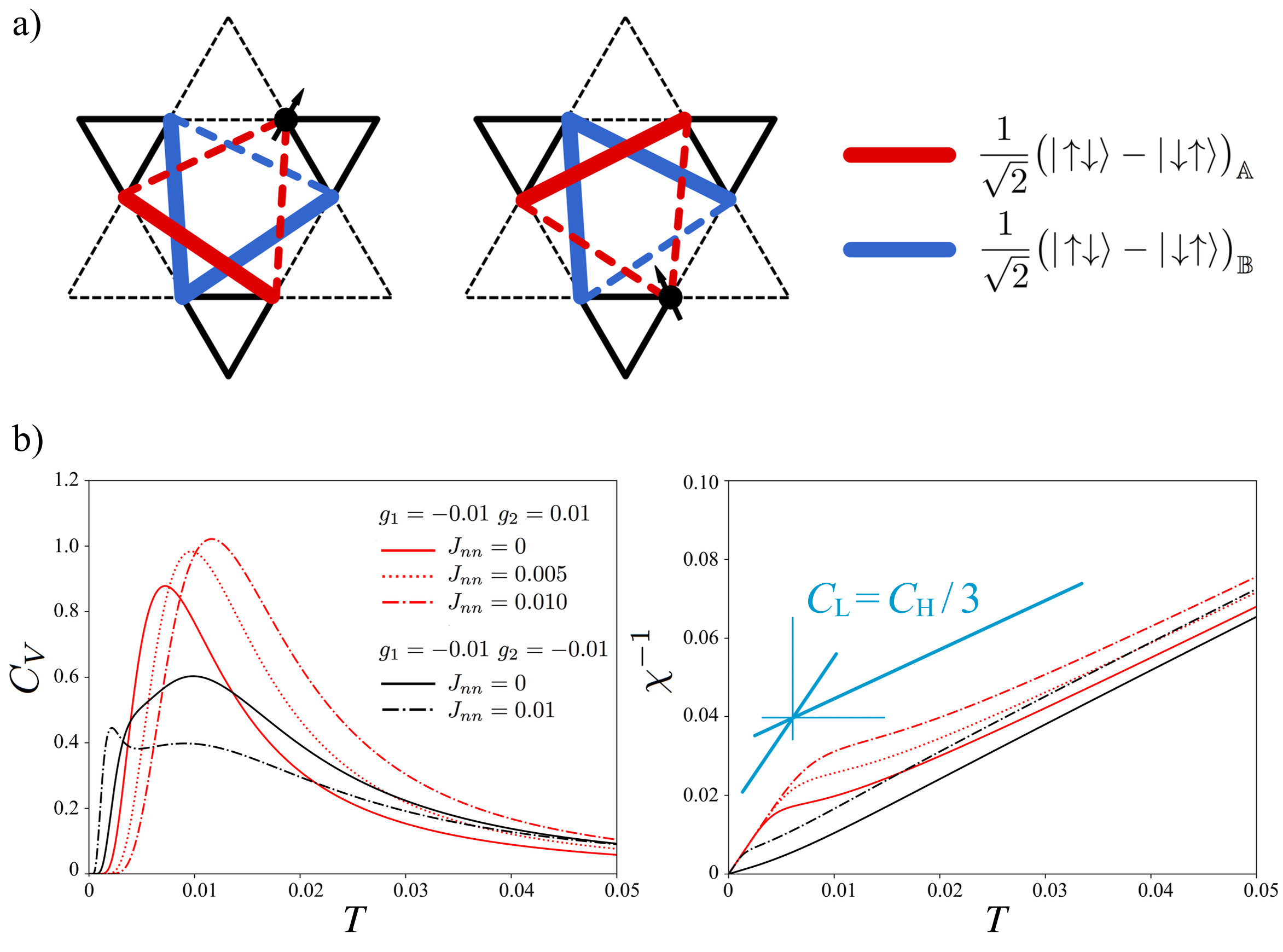}
\end{center}
\caption{a) Valence bonds at the resonating hexagon with one dangling spin; b) Specific heat and inverse spin susceptibility of a single resonating hexagon.}
\label{fig:single}
\end{figure}
%%%%%%%%%%%%%%%%%%%

\par One can further include antiferromagnetic spin fluctuations between next-nearest neighbours in each hexagon $\mathcal{H}_{S}=J_{nn}\sum_{\langle\!\langle ij \rangle\!\rangle}n_{i}n_{j}\big(\boldsymbol{S}_{i}\cdot\boldsymbol{S}_{j}-\frac{1}{4}\big)$, where $J_{nn}=4t^{2}_{nn}/U$ and $t_{nn}$ is the corresponding hopping. Assuming that the PCO effectively decouples hexagons, $\mathcal{H}_{D}=\mathcal{H}_{\hexagon}+\mathcal{H}_{S}$ for a single hexagon can be solved exactly yielding four four-fold degenerate states~\cite{supp}. When $g_{1}$ and $g_{2}$ have opposite signs, regardless of the value of $J_{nn}$ the ground state of $\mathcal{H}_{D}$ displays valence bond condensation with one orphan spin, as shown in Fig.~\ref{fig:single}a:
\begin{equation*}
\begin{aligned}
|\psi_{1}\rangle &= \frac{1}{2}\Big( |\!\uparrow\uparrow\downarrow\rangle_{\mathbb{A}} - |\!\downarrow\uparrow\uparrow\rangle_{\mathbb{A}}  \\
& - \frac{g_{1}-g_{2}}{\tilde{g}} |\!\uparrow\uparrow\downarrow\rangle_{\mathbb{B}} - \frac{g_{2}}{\tilde{g}} |\!\uparrow\downarrow\uparrow\rangle_{\mathbb{B}} + \frac{g_{1}}{\tilde{g}} |\!\downarrow\uparrow\uparrow\rangle_{\mathbb{B}} \Big), \\
|\psi_{2}\rangle &= \frac{1}{2}\Big( |\!\uparrow\downarrow\uparrow\rangle_{\mathbb{A}} - |\!\downarrow\uparrow\uparrow\rangle_{\mathbb{A}} \\
&+ \frac{g_{2}}{\tilde{g}} |\!\uparrow\uparrow\downarrow\rangle_{\mathbb{B}} - \frac{g_{1}}{\tilde{g}} |\!\uparrow\downarrow\uparrow\rangle_{\mathbb{B}} + \frac{g_{1}-g_{2}}{\tilde{g}} |\!\downarrow\uparrow\uparrow\rangle_{\mathbb{B}} \Big).
\end{aligned}
\end{equation*}
\noindent with $\tilde{g}=\sqrt{g_{1}^{2}-g_{1}^{}g_{2}^{}+g_{2}^{2}}$. Such an unusual entanglement with dangling spins originates solely from the asymmetry of tunnelling processes that in turn maximises singlet pairing between the resonating electrons, while the unpaired spins behave paramagnetically in a thermodynamic limit. Interestingly, a similar situation can be realised when $g_{1}<0$ and $g_{2}<0$ with large antiferromagnetic coupling $J_{nn}>\frac{2}{3}(-g_{1}-g_{2}-\tilde{g})$, which was earlier suggested to pair 2/3 of the spins at low temperatures~\cite{chen,chen2}. However, the calculated thermodynamic properties shown in Fig.~\ref{fig:single}b clearly demonstrate that two paramagnetic regimes possess a much higher $T_{C}$ when $g_{1}$ and $g_{2}$ have opposite signs. Our first-principles calculations will show that the strong interaction limit is realised in LiZn$_{2}$Mo$_{3}$O$_{8}$, where $g_{1}$ and $g_{2}$ have opposite signs and $J_{nn}$ is negligibly small.

%%%%%%%%%%%%%%%%%%%
\begin{figure*}[t!]
\begin{center}
\includegraphics[width=0.96\textwidth]{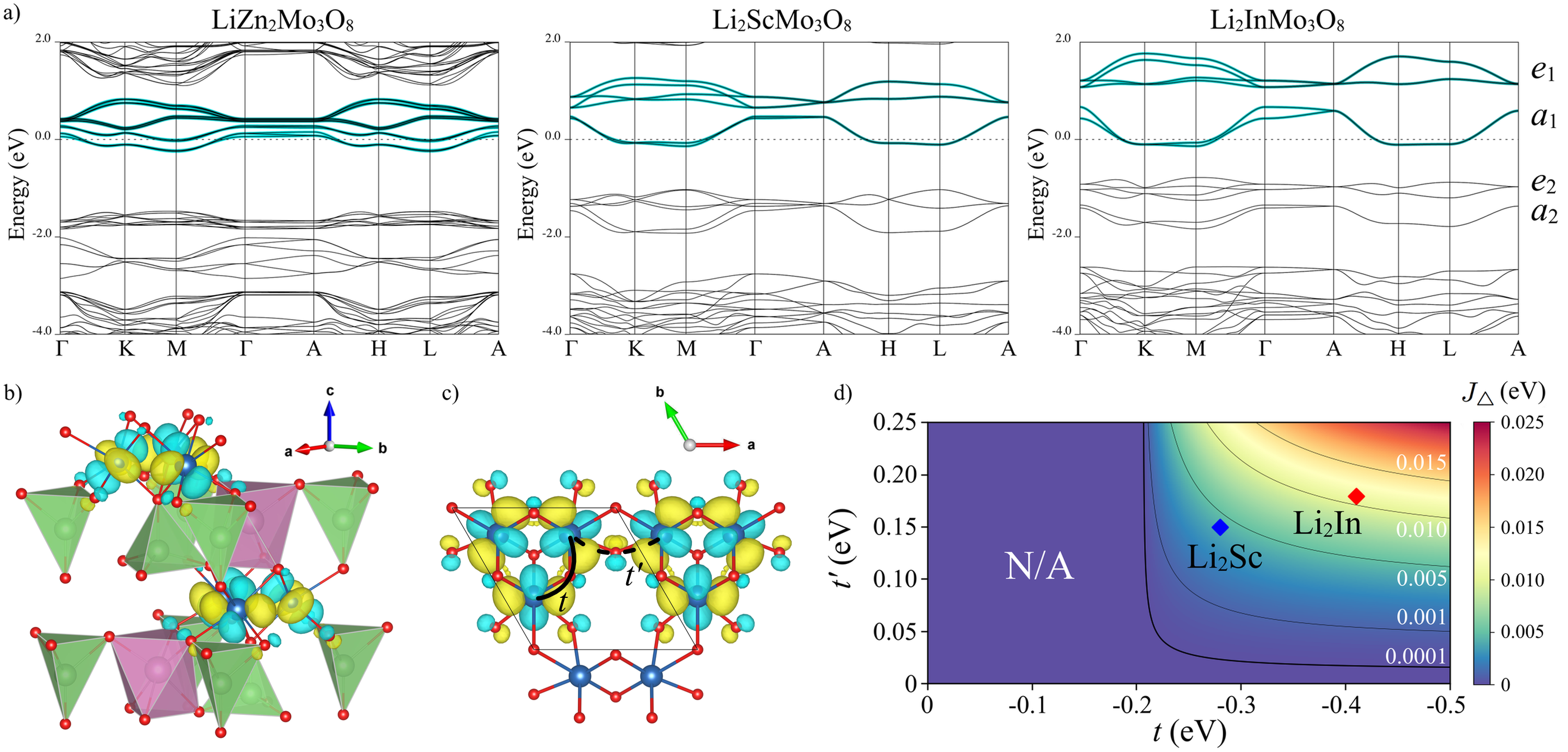}
\end{center}
\caption{a) Band structures of LiZn$_{2}$Mo$_{3}$O$_{8}$, Li$_{2}$ScMo$_{3}$O$_{8}$, and Li$_{2}$InMo$_{3}$O$_{8}$; b) Wannier functions corresponding to the $a_{1}$ and $e_{1}$ states in Li$_{2}$InMo$_{3}$O$_{8}$; c) Wannier functions of the neighbouring Mo$_{3}$O$_{13}$ clusters in one layer of Li$_{2}$InMo$_{3}$O$_{8}$; d) Exchange coupling $J_{\triangle}$ calculated from Eq.~(\ref{eq:clusterspin}) with $U=2.0$~eV, $V=1.1$~eV, and $V'=0.9$~eV. Li$_{2}$ScMo$_{3}$O$_{8}$ and Li$_{2}$InMo$_{3}$O$_{8}$ are schematically shown with diamonds.}
\label{fig:bands}
\end{figure*}
%%%%%%%%%%%%%%%%%%%

\par \emph{Cluster Hubbard Model}.  As $|t|$ increases, the electrons start moving freely within the $\mathcal{T}$ triangle, and the number of electrons at the adjacent $\mathcal{T}'$ triangles fluctuates. When $|t|\sim V'$, the perturbation theory considered above breaks down, and the electrons minimise their energy by forming  bound ``molecular" states. As a result,  the kagom\'e lattice is trimerised, and the original model in Eq.~(\ref{eq:hubsingle}) can be reformulated as a three-orbital extended Hubbard model on the triangular lattice formed by the $\mathcal{T}$ triangles:
\begin{gather*}
\mathcal{H}_{\mathrm{CF}} = \frac{\Delta}{3}\sum_{mm'\in\mathcal{T},\sigma}c^{\dagger\sigma}_{im}\left( \begin{array}{ccc}
0 & 1 & 1 \\
1& 0 & 1 \\
1 &1 & 0 \\
\end{array} \right)_{mm'}\!\!\!\!\!\!\!\!c^{\sigma}_{im'} 
\end{gather*}
\noindent with $\Delta = 3t$. As follows, $\mathcal{H}_{\mathrm{CF}}$ has the form of crystal field that splits the electronic states at the $\mathcal{T}$ triangle into the single $a_{1}$ and double degenerate $e_{1}$ states with energy levels $\frac{2\Delta}{3}$ and $-\frac{\Delta}{3}$, respectively: $|a_{1}\rangle= \frac{1}{\sqrt{3}}\big(|1\rangle + |2\rangle + |3\rangle \big)$, $|e_{1}^{(1)}\rangle= \frac{1}{\sqrt{3}}\big(w|1\rangle + \bar{w}|2\rangle  +  |3\rangle\big)$, and $|e_{1}^{(2)}\rangle= \frac{1}{\sqrt{3}}\big(\bar{w}|1\rangle + w|2\rangle + |3\rangle \big)$ with $\omega=e^{2\pi\mathrm{i}/3}$. Importantly, the $a_{1}$ state is occupied when $\Delta<0$ ($t<0$). Despite the weak interaction limit, the electrons are localised at the $\mathcal{T}$ triangles by their kinetic energy due to a dilute 1/6 filling. We refer to this state as a cluster Mott insulator as opposed to the PCO phase where localisation is entirely driven by intersite Coulomb interactions.
\par As shown in Fig.~\ref{fig:ed}, when both $t<0$ and $t'>0$ are large the localised electrons can develop long-range magnetic order. In this limit, the on-site $\widetilde{U}=\frac{U+2V}{3}$ comes back into play and forbids any double occupancy at the $\mathcal{T}$ triangles, and the corresponding spin model $\mathcal{H}_{\triangle}=\sum_{\langle ij \rangle}J_{\triangle}\boldsymbol{S}_{i}\cdot\boldsymbol{S}_{j}$ on the triangular lattice can be derived to second order in $t'/U$ and $t'/\Delta$:
\begin{equation}
\begin{aligned}
J_{\triangle}=&-\frac{8t'^{2}}{3(2V+3|\Delta|-2V')} + \frac{4t'^{2}}{3(U+2V-2V')} \\
& + \frac{8t'^{2}}{3(U+2V+3|\Delta|-2V')},
\end{aligned}
\label{eq:clusterspin}
\end{equation}
\noindent which can be both ferro- and antiferromagnetic, that explains why some of the recently found Mo$_{3}$O$_{8}$ systems are ferromagnetic insulators~\cite{thesis}. Stability of the magnetic order is directly related to the strength of $t$ and $t'$ in the sense that it can be suppressed by thermal or quantum fluctuations when $t$ or $t'$ are not strong enough to avoid electron fluctuations at the $\mathcal{T'}$ triangles.
 
%%%%%%%%%%%%%%%%%%%%%%%%%%%%
\begin{table}[b]
\caption{Model parameters (in eV) for the one-orbital extended Hubbard model, Eq.~(\ref{eq:hubsingle}). }
\begin{center}
\begin{tabular}{c|c|cc|cc|c}
\hline
\hline
& $U$ & $t$ & $V$ & $t'$ & $V'$ & $t_{nn}$ \\
\hline
LiZn$_{2}$Mo$_{3}$O$_{8}$ & \,\,2.0\, & \,$-0.134\,$ & 0.8\,\, &\,\, 0.113\,\, & 0.6\, & \,0.026 \\
Li$_{2}$ScMo$_{3}$O$_{8}$ & \,\,2.0\, & \,$-0.281\,$ & 1.0\,\, &\,\, 0.147\,\, & 0.8\, & \,0.014\\
Li$_{2}$InMo$_{3}$O$_{8}$ & \,\,2.1\, & \,$-0.409\,$ & 1.2\,\, & \,\, 0.181\,\, & 0.9\, & \,0.010\\
\hline
\hline
\end{tabular}
\end{center}
\label{tab:param}
\end{table}
%%%%%%%%%%%%%%%%%%%%%%%%%%%%

%%%%%%%%%%%%%%%%%%%%%%%%%%%%
\par \emph{First-principles}.~Electronic structure calculations for each system have been performed within local density approximation~\cite{lda} by using projected augmented wave formalism~\cite {paw}, as implemented in \texttt{VASP}~\cite{vasp}, and norm-conserving pseudopotentials, as implemented in \texttt{Quantum ESPRESSO}~\cite{qe}. The calculated band structures are shown in Fig.~\ref{fig:bands}a, indicating the $a_{2}$ and $e_{2}$ states below the Fermi level, which are responsible for the Mo-Mo bonding in the Mo$_{3}$O$_{13}$ cluster, and the molecular $a_{1}$ and $e_{1}$ states occupied by unpaired electrons. The latter were adopted for constructing the extended Hubbard model, Eq.~(\ref{eq:hubsingle}), in the basis of Wannier functions, which were obtained with \texttt{wannier90}~\cite{wan90}, as shown in Fig.~\ref{fig:bands}b. The full set of model parameters is given in Table~\ref{tab:param}~\cite{supp}.
\par According to our results, the splitting between the $a_{1}$ and $e_{1}$ states varies significantly within the systems~\cite{comm2}, and the values of $t/V'$ and $t'/V$ point out at different regimes of electron localisation for each system. Furthermore, $t$ and $t'$ always have opposite signs. This is related to the fact that in the Mo$_{3}$O$_{13}$ clusters with short Mo-Mo bonds the direct $d$-$d$ (always negative) hopping dominates, as shown in Fig.~\ref{fig:bands}c. Because this term vanishes rapidly with metal-metal distance ($\sim1/r^{5}$~\cite{Harrison}), the hopping process via common oxygens having the opposite sign starts to dominate between the clusters, and $t'$ turns out to be positive. We believe that the opposite signs of $t$ and $t'$ is a fundamental aspect of the trimerised kagome lattice at 1/6 filling. According to the general Jahn-Teller theorem, the trimerisation should lift the degeneracy of the ground state so that a single electron resides at the $a_{1}$ orbital of the $\mathcal{T}$ triangle forming a one-dimensional representation of the point group, that occurs only when $t<0$ and $t'>0$. 
\par One can see that $t/V'$ and $t'/V$ are small in LiZn$_{2}$Mo$_{3}$O$_{8}$, preventing the electrons from being localised at the molecular states and thus leading to an emergent PCO with unpaired spins at the resonating hexagons. Moreover, a negligibly small $J_{nn}=1.4$~meV eliminates all previously suggested scenarios for decoupling 1/3 of the spins at low temperatures~\cite{chen,chen2,akbari}. In fact, valence bond condensation in LiZn$_{2}$Mo$_{3}$O$_{8}$ is driven solely by the asymmetry of tunnelling processes caused by the formation of the Mo$_{3}$O$_{13}$ clusters. Given $g_{1}=13.5$~meV and $g_{2}=-40.1$ meV, the calculated $T_{C}\sim92.0$~K between two paramagnetic regimes is in excellent agreement with experiments~\cite{sheckel1,supp}. 
\par In contrast, Li$_{2}$ScMo$_{3}$O$_{8}$ and Li$_{2}$InMo$_{3}$O$_{8}$ have larger splittings between the $a_{1}$ and $e_{1}$ states, and the ratio $t/V'$ favours electron localisation at the Mo$_{3}$O$_{13}$ clusters stabilising a cluster Mott insulator phase. Indeed, having the  largest $t/V'$ and $t'/V$, Li$_{2}$InMo$_{3}$O$_{8}$ reveals an antiferromagnetic order with $J_{\triangle}=9.5$~meV (109.8~K) in good agreement with the experimental value of 112~K~\cite{haraguchi1}. On the other hand, $J_{\triangle}=4.0$~meV (46.7~K) in Li$_{2}$ScMo$_{3}$O$_{8}$, being consistent with the experimental value of 67~K, is close to the instability region where $J_{\triangle}$ is small, as clearly seen in Fig.~\ref{fig:bands}c. Consequently, although the electrons tend to localise at the Mo$_{3}$O$_{13}$ clusters, Li$_{2}$ScMo$_{3}$O$_{8}$ is more likely to fall into an intermediate regime, where any long-range magnetic order is suppressed by quantum fluctuations down to low temperatures. Since the number of electrons at the $\mathcal{T'}$ triangles is allowed to fluctuate when $t/V'$ and $t'/V$ are not strong, we conclude that magnetic order in Li$_{2}$ScMo$_{3}$O$_{8}$ is short-range with QSL-like excitations.
\par \emph{Conclusions}. Having considered an extended Hubbard model on the anisotropic kagom\'e lattice at 1/6 filling, we showed that it features two different limits: a plaquette charge order with one orphan spin as realised in quantum paramagnet LiZn$_{2}$Mo$_{3}$O$_{8}$, and a cluster Mott insulator as revealed in Li$_{2}$InMo$_{3}$O$_{8}$ with a N\'eel-type antiferromagnetic order  and Li$_{2}$ScMo$_{3}$O$_{8}$ with a quantum spin liquid behaviour. Based on first-principles calculations, we demonstrated that their manifestation can be attributed to the trimerisation of the kagom\'e lattice specifying the character of electron localisation, that unravels a largely speculated origin of magnetism in these systems. 
\par Finally, it is known that spin-$\frac{1}{2}$ systems with an odd number of electrons can reveal both long-range order and short-range correlations with topological excitations~\cite{hastings}. While LiZn$_{2}$Mo$_{3}$O$_{8}$ remains a unique example featuring two paramagnetic regimes with unpaired spins, different scenarios of a cluster Mott insulator phase can be realised in other trimerised cluster systems, such as Li$_{2}$In$_{1-x}$Sc$_{x}$Mo$_{3}$O$_{8}$~\cite{akbari}, ScZnMo$_{3}$O$_{8}$~\cite{last1}, and Nb$_{3}$Cl$_{8}$~\cite{last2,last3}.
\par \emph{Acknowledgements}.~S.A.N. thanks Dr.~Wei Ren for stimulating discussions.~I.V.S. and S.V.S.~were supported by projects RFBR 20-32-70019, programs AAAA-A18-118020190095-4 (Quantum) and contract No. 02.A03.21.0006. S.V.S. is grateful for discussions to V.~Irkhin, Yu.~Skryabin, and D.I. Khomskii.

%%%%%%%%%%%%%%%%%%%%%%%%%%%%
%%%%%%%%%%%%%%%%%%%%%%%%%%%%
%%%%%%%%%%%%%%%%%%%%%%%%%%%%


\begin{thebibliography}{99}

\bibitem{moessner}
R.~Moessner, Can. J. Phys. \textbf{79}, 1283 (2001).

\bibitem{patrick1}
Patrick. A.~Lee, Science \textbf{321}, 1306 (2008). 

\bibitem{balents}
L.~Balents, Nature (London) \textbf{464}, 199 (2010).

\bibitem{review}
Lucile Savary, Leon Balents, Rep. Prog. Phys. \textbf{80}, 016502 (2017).

\bibitem{spinliq1} %Spin-liquid phase of the multiple-spin exchange hamiltonian on the triangular lattice
 G. Misguich, C. Lhuillier, B. Bernu, and C. Waldtmann, Phys. Rev. B \textbf{60}, 1064 (1999).

\bibitem{spinliq2} %Variational study of triangular lattice spin-12 model with ring exchanges and spin liquid state in ?-(et)2cu2(cn)3
Olexei I. Motrunich, Phys. Rev. B \textbf{72}, 045105 (2005).

\bibitem{spinliq3} %Chiral spin liquid and emergent anyons in a kagome lattice mott insulator
B. Bauer, L. Cincio, B. P. Keller, M. Dolfi, G. Vidal, S. Trebst, and A. W. W. Ludwig, Nat. Commun. \textbf{5}, 5137 (2014).

\bibitem{spinliq4} %Emergent chiral spin liquid: fractional quantum hall effect in a kagome heisenberg model
Shou-Shu Gong, Wei Zhu, and D. N. Sheng, Scientific reports \textbf{4}, 6317 (2014).

\bibitem{spinliq5} %Quantum Spin Liquid States
Yi Zhou, Kazushi Kanoda, Tai-Kai Ng, Rev. Mod. Phys. \textbf{89}, 025003 (2017).

\bibitem{vorbot}
Z. Hiroi, M. Hanawa, N. Kobayashi, M. Nohara, and H. Takagi, J. Phys. Soc. Jpn. \textbf{70}, 3377 (2001).

\bibitem{herb1}
J. S. Helton \emph{et al.}, Phys. Rev. Lett. \textbf{98}, 107204 (2007).

\bibitem{herb2}
A. Olariu, P. Mendels, F. Bert, F. Duc, J. C. Trombe, M. A. de Vries, and A. Harrison, Phys. Rev. Lett. \textbf{100}, 087202 (2008).

\bibitem{bedt}
Y. Shimizu, K. Miyagawa, K. Kanoda, M. Maesato, and G. Saito, Phys. Rev. Lett. \textbf{91}, 107001 (2003).

\bibitem{dmit1}
T. Itou, A. Oyamada, S. Maegawa, M. Tamura, R. Kato, J. Phys. Conf. Ser. \textbf{145}, 012039 (2009).

\bibitem{dmit2}
T. Itou, A. Oyamada, S. Maegawa, M. Tamura, R. Kato, Phys. Rev. B \textbf{77}, 104413 (2008).

\bibitem{carroll}
W.~H.~McCarroll, Inorg. Chem. \textbf{16}, 3351 (1977).

\bibitem{cotton}
F. A. Cotton, Inorg. Chem. \textbf{3}, 1217 (1964).

\bibitem{sheckel1}
J. P. Sheckelton, J. R. Neilson, D. G. Soltan and T. M. McQueen, Nat. Mater. \textbf{11}, 493 (2012).

\bibitem{sheckel2}
M. Mourigal, W. T. Fuhrman, J. P. Sheckelton, A. Wartelle, J. A. Rodriguez-Rivera, D. L. Abernathy, T. M. McQueen, and C. L. Broholm, Phys. Rev. Lett. \textbf{112}, 027202  (2014).

\bibitem{sheckel3}
J. P. Sheckelton \emph{et al.}, Phys. Rev. B \textbf{89}, 064407 (2014).

\bibitem{flint}
Rebecca Flint and Patrick A.~Lee, Phys. Rev. Lett. \textbf{111}, 217201 (2013).

\bibitem{chen}
Gang Chen, Hae-Young Kee, and Yong Baek Kim, Phys. Rev.~B \textbf{93}, 245134 (2016).

\bibitem{haraguchi1}
Yuya Haraguchi, Chishiro Michioka, Masaki Imai, Hiroaki Ueda, and Kazuyoshi Yoshimura, Phys. Rev.~B \textbf{92}, 014409 (2015).

\bibitem{haraguchi2}
Kazuki Iida \emph{et al.}, Sci. Rep. \textbf{9}, 1826 (2019). 

\bibitem{chen2}
Gang Chen and Patrick A. Lee, Phys. Rev. B \textbf{97}, 035124 (2018).

\bibitem{vesta}
K. Momma and F. Izumi, J.~Appl. Crystallogr. \textbf{44}, 1272 (2011).

\bibitem{supp}
See Supplemental Material at [] for the full set of model parameters and calculation details, which includes Refs.~\cite{mp,wanapp,rpa1,rpa2,interplay}.

\bibitem{thesis}
R. P. Sinclair, PhD Thesis, University of Tennessee, 2018.

\bibitem{dimer1}
R.~Moessner, S.~L.~Sondhi,  and P.~Chandra, Phys. Rev.~B \textbf{64}, 144416 (2001).

\bibitem{dimer2}
Thiago M.~Schlittler, R\'emy Mosseri, and Thomas Barthel, Phys. Rev.~B \textbf{96} 195142 (2017).

\bibitem{kinetic}
F. Pollmann, P. Fulde, and K. Shtengel, Phys. Rev. Lett. \textbf{100}, 136404 (2008).

\bibitem{nishimoto}
Satoshi Nishimoto, Masaaki Nakamura, Aroon O'Brien, and Peter Fulde, Phys. Rev. Lett. \textbf{104}, 196401 (2010).

\bibitem{brien}
A. O'Brien, F. Pollmann, P. Fulde, Phys. Rev. B \textbf{81}, 235115 (2010).

\bibitem{comm1}
Here, $|\psi_{3}\rangle$ and $|\psi_{4}\rangle$ are obtained by applying time-reversal symmetry to the above states. 

\bibitem{lda}
W.~Kohn, L.~J.~Sham, Phys. Rev. A \textbf{140}, 1133 (1965).

\bibitem{paw}
P.~E.~Blochl, Phys. Rev. B \textbf{50}, 17953 (1994).

\bibitem{vasp}
G. Kresse, J. Hafner, Phys. Rev. B \textbf{47}, 558 (1993).

\bibitem{qe}
P.~Giannozzi \emph{et al.}, J.~Phys.: Condens.Matter \textbf{21}, 395502 (2009).

\bibitem{wan90}
A.~A.~Mostofi, J.~R.~Yates, G.~Pizzi, Y.~S.~Lee, I.~Souza, D.~Vanderbilt, N.~Marzari, Comput. Phys. Commun. \textbf{185}, 2309 (2014).

\bibitem{comm2}
In LiZn$_{2}$Mo$_{3}$O$_{8}$, there is a small splitting between the $a_{1}$ and $e_{1}$ states corresponding to different Mo layers due to disorder of Li and Zn ions.

\bibitem{Harrison} 
W.~A.~Harrison, Elementary Electronic Structure (World Scientific, Singapore, 1999).

\bibitem{akbari}
A. Akbari-Sharbaf, R. Sinclair, A. Verrier, D. Ziat, H. D. Zhou, X. F. Sun, and J. A. Quilliam, Phys. Rev. Lett. \textbf{120}, 227201 (2018).

\bibitem{hastings}
M. B. Hastings, Phys. Rev. B \textbf{69}, 104431 (2004).

\bibitem{last1}  %Zn3Mo3O8, and ScZnMo3O8
C. C. Torardi, R. E. McCarley, Inorg. Chem. ,\textbf{24}, 476 (1985).

\bibitem{last2} %Nb3Cl8
Yuya Haraguchi, Chishiro Michioka, Manabu Ishikawa, Yoshiaki Nakano, Hideki Yamochi, Hiroaki Ueda, Kazuyoshi Yoshimura, Inorg. Chem. 56, 3483 (2017).

\bibitem{last3} 
J. P. Sheckelton, K. W. Plumb, B. A. Trump, C. L. Broholm, and T. M. McQueen, Inorg. Chem. Front. \textbf{4}, 481 (2017).

%%%% SM
\bibitem{mp}
H.~J.~Monkhorst, J.~D.~Pack, Phys. Rev. B 13, \textbf{5188} (1976).

\bibitem{wanapp}
N.~Marzari, A.~A.~Mostofi, J.~R.~Yates, I.~Souza, and D.~Vanderbilt, Rev. Mod. Phys. \textbf{84}, 1419 (2012).

\bibitem{rpa1}
M.~Springer and F.~Aryasetiawan, Phys. Rev. B \textbf{57}, 4364 (1998).

\bibitem{rpa2}
F.~Aryasetiawan, M.~Imada, A.~Georges, G.~Kotliar, S.~Biermann, and A.~I.~Lichtenstein, Phys. Rev. B \textbf{70}, 195104 (2004).

\bibitem{interplay}
Frank Pollmann, Krishanu Roychowdhury, Chisa Hotta, and Karlo Penc, Phys. Rev.~B \textbf{90}, 035118 (2014).

\end{thebibliography}
\end{document}